\documentclass[preprint,12pt]{elsarticle}

\usepackage{amssymb}
\usepackage{amsmath}
\usepackage{graphicx}
\usepackage{caption}  
\usepackage{tabularx}
\usepackage{float}  
\usepackage{url}
\usepackage{sidecap}
\usepackage{diagbox}
\usepackage{array}
\usepackage{multirow}
\usepackage{booktabs}
\usepackage{subcaption}
\usepackage{amsmath,amsfonts}
\usepackage{colortbl}
\usepackage{algorithm,algorithmic}
\usepackage{fancyhdr}
\usepackage{bm}
\usepackage{breqn}
\usepackage{xcolor}
\usepackage{soul}
\usepackage{geometry}
\usepackage[breaklinks=true]{hyperref}
\usepackage{orcidlink}

\makeatletter
\def\ps@pprintTitle{%
  \let\@oddhead\@empty
  \let\@evenhead\@empty
  \let\@oddfoot\@empty
  \let\@evenfoot\@empty
}
\makeatother

\date{}
\begin{document}

\begin{frontmatter}



\title{Domain-Adaptive Arrhythmia Classification Using a Hybrid Transformer on Wearable Heart Signals}

\author[label2]{Maedeh H. Toosi\corref{cor1} \orcidlink{https://orcid.org/0009-0006-4710-4853}}
\ead{maedeh.toosi@ut.ac.ir}

\author[label2]{Siamak Mohammadi \orcidlink{https://orcid.org/0000-0003-1515-7281}}
\ead{smohamadi@ut.ac.ir}


\affiliation[label2]{organization={School of Electrical and Computer Engineering, University of Tehran},
            city={Tehran},
            country={Iran}}


\begin{abstract}
Cardiovascular disease remains the leading cause of death globally, underscoring the need for effective, accessible monitoring solutions, particularly through wearable devices that enable continuous, real-time tracking of heart rhythms in home settings. However, deploying deep learning models trained on clinical electrocardiogram (ECG) datasets to wearable devices remains challenging, as differences in recording equipment, signal quality, and patient populations introduce domain shifts that degrade model performance. We propose a hybrid transformer model that processes continuous ECG signals alongside seven heart rate variability (HRV) features, where the raw signal path captures beat-level morphological patterns and the HRV path encodes rhythm regularity statistics, allowing the model to jointly leverage complementary information from both representations. To enhance the model's ability to generalize across domains, we employ representation learning techniques, including Maximum Mean Discrepancy (MMD), a non-parametric kernel-based metric that quantifies the distance between feature distributions of different domains, to align feature distributions between source and target domains, addressing the challenge of domain shifts between public datasets and wearable device data. By leveraging five public ECG datasets for training, the model learns robust, generalized representations that mitigate domain-specific biases. When tested on wearable device data with an unseen domain, the model achieved an F1-macro 95\% and balanced accuracy of 96.15\%. These results demonstrate minimal performance degradation, with only a 2\% drop in F1-macro compared to seen-domain evaluation, highlighting the model's generalization capabilities and its potential for reliable, real-time heart monitoring applications in home and ambulatory settings.
\end{abstract}


\begin{highlights}
\item A hybrid Transformer model is proposed for domain-adaptive arrhythmia classification. 
\item The model synergistically processes raw ECG signals and derived Heart Rate Variability (HRV) features.
\item Domain adaptation using Maximum Mean Discrepancy improves cross-dataset generalization.
\item  Achieved F1-macro of 95\% and balanced accuracy of 96.15\% on wearable ECG signals.
\item  Demonstrates robustness to noise, enabling reliable real-world wearable ECG monitoring.

\end{highlights}


\end{frontmatter}



\section{Introduction}
\label{sec:introduction}
Cardiovascular disease (CVD) remains the leading cause of death globally, requiring ongoing advancements in prevention and treatment strategies~\cite{world2023global}. The need for more effective and accessible monitoring solutions has grown, particularly for arrhythmias, due to the persistent impact of cardiovascular diseases. Significant progress has been made in developing tools to predict CVD events over the past few decades, reducing the burden of this global health challenge\cite{joseph2017reducing,grant2024evolving}. 
Wearable devices play a crucial role in home monitoring for CVD, offering continuous, real-time tracking of vital health parameters such as heart rate and arrhythmias.

Despite their potential, the adoption of wearable devices remains limited, highlighting the need for reliable and generalizable AI models that can work effectively across diverse wearable platforms~\cite{gadaleta2023prediction,biton2023generalizable}.
Despite the benefits that wearable devices offer, there is potential to further enhance their effectiveness through the integration of artificial intelligence (AI). AI can analyze vast amounts of data collected by these devices to detect patterns and predict potential health issues before they become critical~\cite{williams2023wearable}.
However, the implementation of AI in healthcare presents substantial challenges due to variations in medical equipment, the complexities of clinical environments, and the need for large, diverse training datasets~\cite{rajpurkar2022ai,gerke2020ethical}.

Collecting comprehensive medical data is challenging due to the diversity of devices and clinical settings, leading to single-source bias where AI models overfit to specific environments, thus reducing their generalizability across different healthcare settings~\cite {hawkins2004problem}. Moreover, the quality and diversity of medical data are often compromised due to extensive data cleaning and preprocessing. While necessary, these processes can remove real-world complexities, resulting in AI systems that may not accurately reflect clinical realities~\cite{rajpurkar2022ai,park2020artificial}.

\begin{figure*}[t]
\centerline{\includegraphics[width=0.99\linewidth]{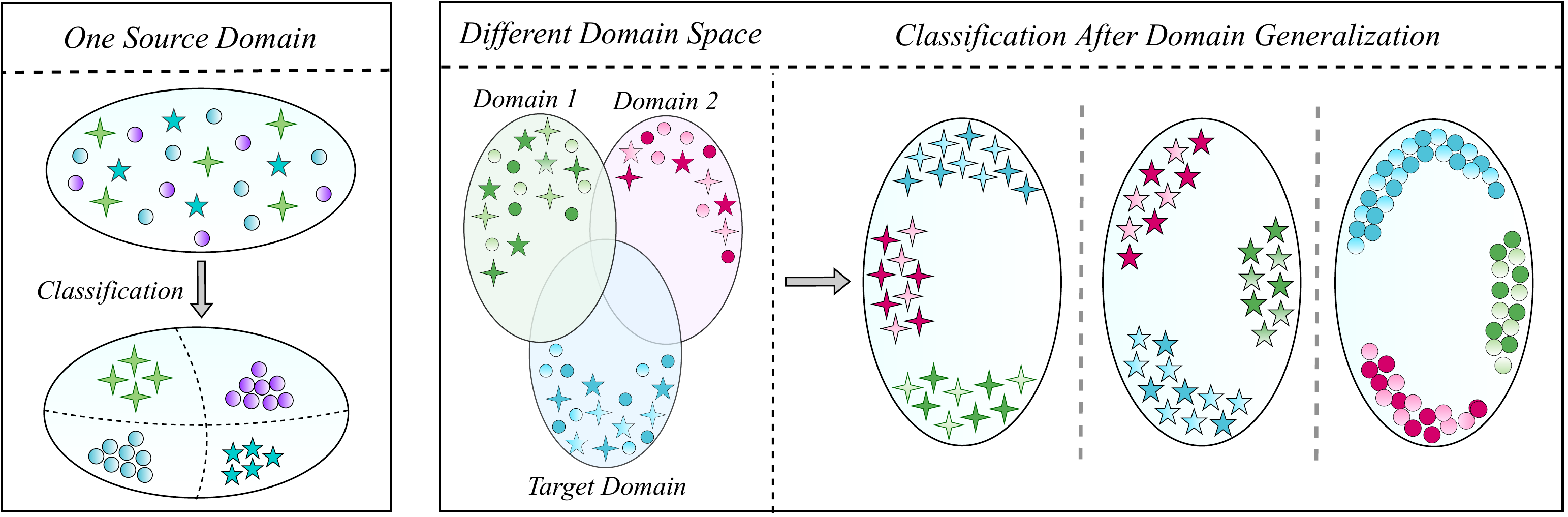}}
\caption{Comparison of Classification Performance Between Source and Target Domains. Different shapes (circles, stars, triangles) represent distinct classes, and colors indicate domain membership. The left panel shows accurate classification within a single source domain. The middle panel illustrates the domain shift problem, where samples from two domains occupy different regions of the feature space, leading to misclassification. The right panel demonstrates classification after domain generalization, where feature distributions from both domains are aligned, enabling correct classification across domains.}
\label{domain}
\end{figure*}

\begin{figure*}[t]
\centering
\begin{subfigure}[b]{.49\textwidth}
  \includegraphics[width=\linewidth]{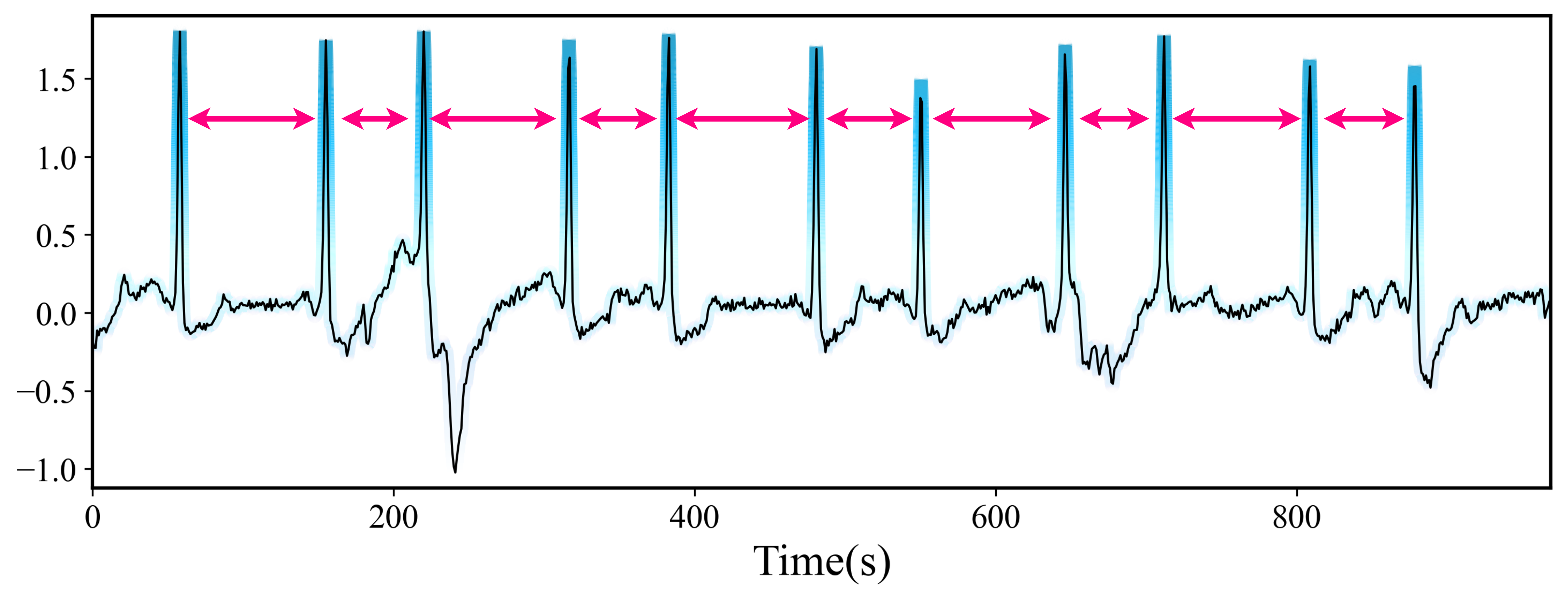}
  \caption{Irregular rhythm with fibrillatory waves during RR-intervals in AF ECG signal}
  \label{af_fig1:a}
\end{subfigure}
\hfill
\begin{subfigure}[b]{.49\textwidth}
  \includegraphics[width=\linewidth]{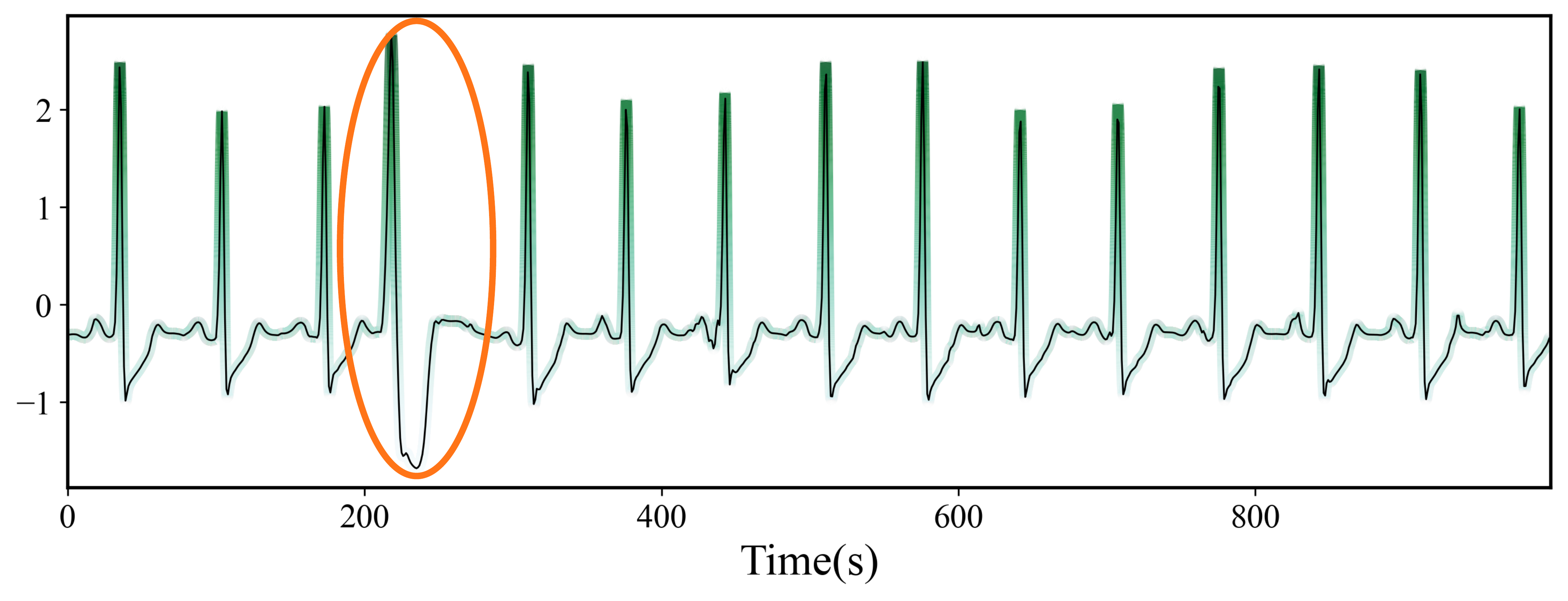}
  \caption{Irregular beats originate in the heart's lower chambers, indicating PVC in the ECG signal}
  \label{Nfig1:c}
\end{subfigure}
\caption{Comparison of ECG Signals for Different Arrhythmias: The left panel shows an atrial fibrillation (AF) signal with irregular RR intervals, while the right panel illustrates a PVC signal, marked by an abnormal beat (highlighted in orange).}
\label{signal}
\end{figure*}

In this context, domain adaptation emerges as a crucial technique in machine learning, addressing the challenge of applying models trained on one domain (source) to another (target) domain, where data characteristics may differ significantly. These domain shifts can lead to performance degradation, as models often struggle to generalize to new environments. Figure~\ref{domain} illustrates this challenge, showing the difference in model classification when trained on a single source domain compared to the difficulties encountered in a different target domain. Such discrepancies highlight the need for methods that enhance a model's ability to adapt across diverse domains, especially in healthcare, where variations in medical data across devices and clinical settings are common\cite{lasko2024probabilistic,bachtiger2022point,ben2010theory}.
Beyond the challenges of domain generalization and data limitations, another key challenge we encounter in this study is the inherent complexity of ECG signal analysis, particularly for Atrial Fibrillation (AF) and Premature Ventricular Contraction(PVC) detection, which is the focus of our case study. As illustrated in Figure~\ref{signal}, AF detection relies on analyzing sustained irregularities in RR intervals over longer durations, focusing on temporal patterns indicative of irregular rhythms. Conversely, PVC detection focuses on beat-level features, examining individual QRS complexes for morphological abnormalities and premature beats~\cite{marcus2020evaluation,developed2010guidelines}.

Recent studies have addressed domain challenges using several methods, including Few-Shot learning ~\cite{oh2022understanding,song2022efficient,ng2023few,amirshahi2024metawears}, learn latent feature across domains with Generative Adversarial Network(GAN) ~\cite{li2018domain,ganin2016domain,tzeng2017adversarial}, Self-Supervised Learning ~\cite{lai2023practical,yue2021prototypical,achituve2021self,ragab2022self}, as well as other techniques like data augmentation and style transfer ~\cite{carlucci2019domain}, domain randomization~\cite{tobin2017domain}.

In this paper, we propose a novel hybrid transformer model designed to improve cross-domain generalization in ECG signal analysis. Our model processes a continuous ECG signal through one input path while simultaneously incorporating seven HRV features calculated from RR intervals in another path. This dual-path approach enables the model to capture both the raw signal dynamics and the physiological significance encapsulated in HRV features. To further mitigate the effects of domain shift, we integrate Maximum Mean Discrepancy (MMD)~\cite{gretton2012optimal} into our training process, aiming to align the feature distributions between the public datasets and our proprietary wearable device data. In addition to domain generalization, our model addresses the challenge of detecting different types of arrhythmias that require distinct analytical methods. By combining raw ECG signals with HRV features, our hybrid transformer model effectively detects both AF and PVC within a unified framework.

Given the limited amount of data available from our wearable device, we leverage five publicly available ECG datasets for training. These datasets, collected from diverse sources and populations, provide a wide variety of ECG patterns and characteristics. By training on multiple datasets, our model learns generalized representations that are more robust to domain variations. Testing on our limited wearable device data then serves as a practical evaluation of the model's ability to generalize to new, unseen domains.

The main contributions of this work are as follows:
\begin{itemize}

\item We have developed a hybrid transformer model that effectively combines raw ECG signals and HRV features to improve the detection of both AF and PVC arrhythmias.
\item We have employed a supervised representation learning approach augmented with Maximum Mean Discrepancy (MMD)-based unsupervised domain adaptation. Although our dataset has labels, we leverage representation learning techniques to minimize the domain distribution mismatch, thus enhancing generalization
\item We have demonstrated that training on multiple public datasets can compensate for limited proprietary data, providing a viable solution for scenarios where data collection is challenging.
\end{itemize}

\begin{table*}[h]
\centering
\caption{RR-Interval Features and Formulas}
\label{rr_interval_features}
\renewcommand{\arraystretch}{1.7} 
\small 
\resizebox{\textwidth}{!}{
\begin{tabularx}{\textwidth}{|p{2.5cm}|X|X|}
\hline
\textbf{Feature Name} & \textbf{Definition} & \textbf{Formula} \\ \hline
Mean & Average of RR intervals & $\mu = \frac{1}{N} \sum_{i=1}^{N} RR_i$ \\ \hline
STD & Standard deviation of RR intervals & $\sigma = \sqrt{\frac{1}{N} \sum_{i=1}^{N} (RR_i - \mu)^2}$ \\ \hline
Shannon Entropy & Measure of uncertainty in RR interval distribution & $H = -\sum p_i \log p_i$ \\ \hline
RMSSD & Root mean square of successive differences between RR intervals & $\text{RMSSD} = \sqrt{\frac{1}{N-1} \sum_{i=1}^{N-1} (RR_{i+1} - RR_i)^2}$ \\ \hline
Normalized RMSSD & 	RMSSD (Root Mean Square of Successive Differences) divided by the mean of RR intervals & $\text{nRMSSD} = \frac{\text{RMSSD}}{\mu}$ \\ \hline
Mean Absolute Deviation & Mean of absolute deviations from the mean of RR intervals & $\text{MeanAD} = \frac{1}{N} \sum_{i=1}^{N} |RR_i - \mu|$ \\ \hline
Median Absolute Deviation & Median of absolute deviations from the median of RR intervals & $\text{MedianAD} = \text{median}(|RR_i - \text{median}(RR)|)$ \\ \hline
\end{tabularx}
}
\end{table*}
\section{Related Works}
\textbf{Cross-Domain} challenges in the field of medical applications are critical and need to be addressed. This involves leveraging diverse datasets to enhance model robustness and applicability in cross-domain scenarios. Previous studies \cite{zhou2023ensemble,placido2023deep,zhang2021deep} showcase research in this domain, exploring various methods suitable for achieving this goal.
Additionally, recent studies \cite{gadaleta2023prediction,arnaout2021ensemble,wang2021inter,chen2020unsupervised}, further exemplify the methodologies discussed in the context of arrhythmia classification, investigating specific challenges and solutions in this area.
 However, a common limitation is the limited availability of training samples.
This scarcity of data necessitates innovative approaches to training models effectively with a limited number of data. 

\textbf{Few-Shot Learning} has seen recent advancements that present a promising solution to the issue of limited data availability in medical applications.
As demonstrated in \cite{zhu2022enhancing,wang2022high,cui2020unified,liu2021metaphys}, research in this area has further explored its versatility and effectiveness across various medical applications.
In the specific field of ECG classification, Siamese networks have been effectively applied to tackle data scarcity. The study in \cite{gupta2021similarity} utilized Siamese Convolutional Neural Networks, outperforming traditional methods like Dynamic Time Warping and LSTM-FCN. Another study, \cite{li2021one}, employed one-dimensional Siamese networks for ECG diagnosis, focusing on feature extraction via shared-weight CNNs and K-nearest similarity judgment for model testing. Additionally, \cite{chen2022meta} combined meta-learning and transfer learning to expedite learning in ECG arrhythmia detection with limited data availability.

\textbf{Transformer} have significantly impacted the medical field,  enhancing the predictive power and interpretability of deep learning algorithms across various tasks, including medical image classification \cite{ai2020correlation,lu2021smile,shome2021covid} and segmentation\cite{chen2021transunet,hatamizadeh2022unetr,karimi2021convolution}. Their capacity to recognize complex patterns in time series data is particularly important in the healthcare industry; this is demonstrated by their application in the analysis of ECG signals to identify and classify arrhythmias. To identify arrhythmias from ECG signals effectively, authors in~\cite{hu2022transformer} develop a vision transformer model named ECG DETR  with a dilated residual network. The DiResViT model is introduced by \cite{pratiher2022dilated}, which combines dilated convolutions with vision transformers to detect atrial fibrillation using time-frequency representations of ECG signals. For enhanced ECG signal analysis and arrhythmia classification, authors in~\cite{che2021constrained} combine CNNs and a transformer model with a unique link constraint.

\section{Methodology}
\label{proposed_method}
\begin{figure*}[t]
\centerline{\includegraphics[width=0.99\linewidth,height = 5cm]{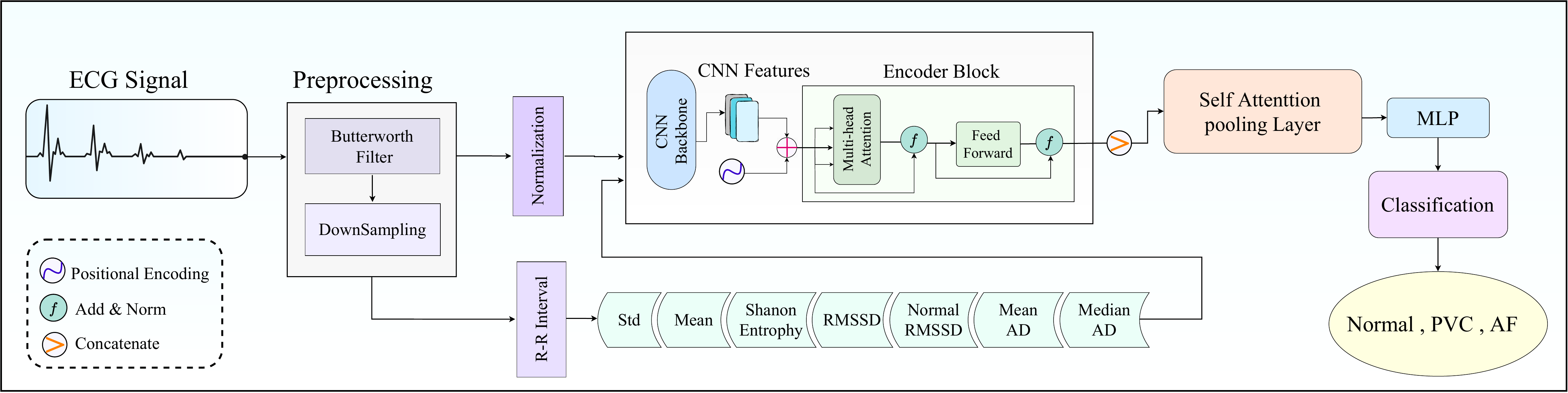}}
\caption{Overview of the proposed model, a comprehensive illustration of the proposed transformer-based model for ECG signal classification. The architecture includes a CNN backbone for feature extraction, a series of transformer encoder layers for deep analysis, and fully connected layers for final classification.}
\label{model-architecture}
\end{figure*}

\subsection{Data Pre-processing}

In this study, the ECG data were collected from various sources, which required preprocessing steps to improve signal quality and ensure consistency across datasets. The following preprocessing techniques were applied:

\textbf{Median Filtering for Baseline Wander Removal:}
Initially, a median filter was applied to remove baseline wander, which is a low-frequency artifact often caused by respiration or patient movement. This filtering step effectively reduced the baseline noise without distorting the ECG signal’s morphology.

\textbf{Butterworth Filter for Noise Elimination:}
To further enhance signal quality, a second-order band-pass Butterworth filter was applied to each ECG recording. This filter was designed to eliminate both baseline drift and high-frequency noise, typically caused by electrical interference or muscle activity.

\textbf{Resampling for Uniformity:}
Given that the ECG signals were recorded at various sampling rates, ranging from 128 Hz to 500 Hz, resampling was necessary to ensure consistency across datasets. All signals were resampled to a uniform rate of 100 Hz. This resampling process ensures that all data have the same resolution, which is crucial when combining datasets from different sources.

\textbf {Segmentation:}
After resampling, the ECG signals were segmented into consecutive 10-second windows, each containing 1000 samples at a 100 Hz sampling rate. The 10-second window was selected to balance capturing meaningful cardiac dynamics with maintaining a manageable input size for machine learning. Prior work has shown that short ECG segments, including 10-second recordings, are effective for cardiovascular event prediction and arrhythmia detection, supporting their use for rapid, low-cost, and scalable analysis~\cite{attia2019artificial,sangha2022automated,khurshid2022ecg,biton2021atrial}.

\textbf{Normalization:}
Finally, each ECG signal was normalized to a range between -1 and 1 to facilitate consistent analysis across all datasets. This normalization step reduces the potential impact of amplitude variations caused by different recording equipment.

\subsection{Feature extraction}
We address the challenges of detecting multiple arrhythmias within the same ECG window by designing a hybrid model that incorporates both beat morphology and inter-beat interval variability. We extracted seven time-domain heart rate variability (HRV) features from the segmented RR intervals (RRI) to capture rhythm dynamics. Although traditional HRV analysis often uses longer recordings, recent studies have demonstrated that HRV features extracted from short ECG segments, including 10-second recordings, are effective for arrhythmia detection and cardiovascular event prediction~\cite{clifford2017af,wan2024novel,jahan2022short}.These features, essential for evaluating heart rhythm variability, enhance our ability to distinguish among normal sinus rhythm, AF, and PVC. Table~\ref{rr_interval_features} provides the definitions and formulas of the selected features. R peaks from all ECG records are extracted using the PhysioNet WFDB Toolbox~\cite{che2021constrained,xie2022waveform}, which applies the Pan–Tompkins~\cite{pan1985real} algorithm for detection.

\subsection{Model Architecture}

Our proposed model, illustrated in Figure \ref{model-architecture}, processes continuous ECG signals to detect arrhythmia using both raw ECG data and extracted R-R interval features. We chose this dual-path approach because different heart issues require different viewpoints: while the specific shape of a heartbeat helps identify PVC, the overall rhythm pattern is necessary to detect AF. By combining these two paths, our model gains a more complete physiological understanding than models that look at the signal alone. The originality of this architecture lies in using a Transformer to balance raw data with statistical insights, which makes the system much more reliable for wearable devices where signals are often noisy or inconsistent.
Specifically, the architecture consists of four main parts: 

(1) CNN Backbones: These layers act as a feature extractor to identify the local "shape" or morphology of the ECG beats.
(2) RR Interval Path: This separate pipeline calculates statistical features from the timing between beats to capture rhythm dynamics.
(3) Transformer Encoders: These layers use an attention mechanism to find complex, long-distance relationships between the signals and the heart rate stats.
(4) Classification Module: This final stage uses fully connected layers to calculate the probability of a patient having a normal rhythm, PVC, or AF.

\textbf{CNN Backbones:} 

 

The CNN backbone plays an important role in extracting local features from ECG signals, taking advantage of the strengths of CNNs in identifying patterns. 
Our ECG CNN architecture consists of three one-dimensional convolutional layers designed to capture a wide range of features from the ECG data. This results in a sequence of embedded representations ($x[0], \ldots, x[n]$). As described in the original transformer paper~\cite{vaswani2017attention}, these embedded representations are added with positional encodings ($p[0], \ldots, p[n]$) to represent the order of each sequence.

In addition to the ECG signals, we incorporate statistical features extracted from the RR intervals to enhance the model's ability to detect predicted output. The RR interval features include statistical measures, as mentioned in Table ~\ref{rr_interval_features}, which provide insights into heart rate variability.

\textbf{RR Interval Features Processing Path:}

The RR interval features are processed through a separate CNN path consisting of two one-dimensional convolutional layers. The first convolutional layer applies 128 filters with a kernel size of 3 to capture local patterns among the statistical features. The second convolutional layer reduces the number of channels to 32 (matching the embedding size $d_{\text{model}}$) with a kernel size of 3. Positional encodings are added to the outputs to retain the sequential information inherent in the features.

\begin{table}[ht]
\centering
\caption{CNN Backbone Settings for Both Processing Paths}
\renewcommand{\arraystretch}{1.3} 
\begin{tabular}{|p{2.6cm}|p{1.9cm}|p{2.2cm}|p{1.4cm}|}
\hline
\textbf{Path} & \textbf{Layer} & \textbf{Channels(In-Out)} & \textbf{K/S/P} \\
\hline
\multirow{5}{*}{\textit{ECG Signal}} & Conv1 & 1 - 64 & 3 / 1 / 0 \\
& Conv2 & 64 - 128 & 3 / 1 / 1 \\
& Conv3 & 128 - 128 & 3 / 1 / 1 \\
& Conv4 & 128 - 32 & 3 / 1 / 1 \\
& MaxPool1d & - & 2 / 2 / 0 \\
\hline
\multirow{2}{*}{\textit{Features}} & Conv1 & 1 - 128 & 3 / 1 / 0 \\
& Conv2 & 128 - 32 & 3 / 1 / 0 \\
\hline
\end{tabular}
\label{combined_cnn}
\end{table}

\begin{table*}[h]
\centering
\caption{A listing of hyper-parameters selected to train the neural network model for ECG finding classification.}
\renewcommand{\arraystretch}{1.3} 
\label{hyperparameter}
\begin{tabular}{|p{4cm}|p{7cm}|p{2cm}|}  
\hline
\textbf{Section} & \textbf{Hyper-Parameter} & \textbf{Value} \\
\hline
\multirow{4}{*}{\textit{Transformer}}  
& Number of encoding layers  & 4 \\
& Embedding size & 32 \\
& Number of heads & 4 \\
& Dimension of feed-forward layer & 64 \\
& Dropout & 0.25 \\
\hline
\multirow{4}{*}{\textit{Fully Connected Layers}} 
& ECG path decoder output size & 32 \\
& RR interval path decoder output size & 64 \\
& Concatenated FC layer size & 32 \\
& Number of classes & 3 \\
\hline
\multirow{3}{*}{\textit{Training Parameters}}  
& Batch size & 16 \\
& Learning rate & 0.001 \\
& Number of epochs & 50 \\
\hline
\end{tabular}
\end{table*}

\textbf{Transformer Encoder Layers:} 
The positionally encoded embeddings from the ECG CNN backbone and the RR interval CNN path are fed into separate stacks of transformer encoder layers. The transformer encoders capture long-range dependencies and contextual relationships within each input. This allows the model to understand complex patterns associated with AF in both the raw signals and the statistical features.

\textbf{Classification Layer:} 
After the transformer encoder layers, the high-level representations of both inputs are combined. This comprehensive approach combines the detailed temporal dynamics of the ECG signals with the statistical insights from the RR intervals. The resulting embeddings then go through a series of fully connected layers for classification. The model provides the probability of arrhythmia presence for each patient as output.

The details for each convolution operation in both processing paths are listed in Table \ref{combined_cnn}, and the hyper-parameters selected for training are provided in Table \ref{hyperparameter}.

\section{Experiments}
\label{sec:Expriment}
\subsection{Datasets}
All ECG signals used in this study, including public training datasets and wearable device test data, were preprocessed following the procedure described in Section~\ref{proposed_method}. The preprocessing steps included median filtering for baseline wander removal, Butterworth band-pass filtering, resampling to 100 Hz, segmentation into 10-second windows, and normalization to the range [-1, 1]. These steps ensured signal quality and uniformity across all datasets, facilitating effective model training and evaluation.
\subsubsection{Train Datasets}

The proposed model has been trained using five public datasets: MIT-BIH Arrhythmia~\cite{moody2001impact}, MIT-BIH SVDB~\cite{greenwald1990improved}, MIT-BIH AF~\cite{moody1983new}, Large Scale dataset~\cite{zheng2022large}, and INCART. Further details about these datasets are provided in Table ~\ref{train_data}.
While these datasets contain additional ECG classes, we report only the classes relevant to our study—Normal, PVC, and AF. All available samples for the AF and PVC classes were used to capture the full range of variability in these arrhythmias. For the Normal class, we used all available signals from the Large-Scale Dataset, as these samples come from different individuals and offer greater variability. However, to prevent overfitting and maintain a balanced dataset, we randomly selected a smaller subset of Normal samples from the remaining datasets.
Table~\ref{train_data} provides a detailed overview of the number of samples in each dataset.

\begin{table}[h]
\centering
\renewcommand{\arraystretch}{1.2}
\caption{Overview of ECG signal samples used for model training across public datasets}
\label{train_data}
\begin{tabular}{lcccccccc}
\toprule
\multirow{2}{*}{{Database}} & \multicolumn{3}{c}{{Classes}} & \multirow{2}{*}{{Frequency}} & \multirow{2}{*}{{No. of }} \\ 
\cmidrule(lr){2-4}
 & {Normal} & {PVC} & {AF}  &  &  records \\ 
\midrule
\midrule
MIT-BIH  & 89507 & 6972 & --- & 360 & 44 \\ 
SVDB & 162339 & 9943 & --- & 128 & 78   \\ 
AF & 45432 & --- & 25424 & 250 & 23  \\ 
Large Scale & 8125 & 294 & 1780& 500 & 45152 \\ 
INCART & 150410  & 20013 &  --- & 257 &  75 \\ 

\bottomrule
\end{tabular}
\end{table}

\subsubsection{Test Dataset}  
The ECG signals were collected using a custom ultra-low-power wearable device developed by~\cite{hafshejani2021ultra}, which acquires single-lead ECG signals and transmits them to a smartphone via Bluetooth Low Energy (BLE). This device includes onboard motion artifact removal at both analog and digital stages, with data sampled at a frequency of 200 Hz, suitable for arrhythmia analysis. The raw ECG recordings were obtained from collaborators who conducted the data collection, and were provided as raw signal files for offline analysis. The recordings were first labeled by medical experts at the whole-record level into four categories: Normal (94 records), PVC (42 records), AF (43 records), and PVC-AF (9 records, indicating patients with both PVC and AF simultaneously). The normal recordings ranged up to 40 seconds, while other recordings ranged between 40 seconds and 2 hours. We then segmented the signals into 10-second intervals, resulting in 3762 segments for PVC and 445 segments for AF. Each 10-second segment was individually reviewed and labeled by medical experts to ensure the accuracy and consistency of the classification. All preprocessing steps described in Section 3.1, including filtering, resampling to 100 Hz, segmentation, and normalization, were applied consistently to this dataset to ensure compatibility with the training data. The processed and labeled dataset used in this study is publicly available and described in the Data Availability section at the end of the paper.

\subsection{Experimental Setup}

As mentioned in Section ~\ref{sec:introduction}, we are dealing with ECG signals collected from different sources (domains), and we aim to train a model that generalizes well to unseen target domains. To address the challenge of domain shifts in ECG signal analysis, we integrate the Maximum Mean Discrepancy (MMD) into our model's training phase~\cite{gretton2012optimal}.  MMD is a non-parametric kernel-based metric used to measure the distance between the probability distributions of the source and target domains in the feature space.  In our implementation, we compute the squared MMD using a Gaussian (RBF) kernel of the form $
k(x, y)=\exp \left(-\beta\|x-y\|^2\right)
$, which is a positive-definite kernel commonly used in domain adaptation.

MMD is applied dynamically during training, it is computed for each mini-batch of feature representations from the source and target domains. By incorporating the MMD into the loss function, we encourage the learned features to be domain-invariant, thus reducing the discrepancy between source and target distributions in latent space. The total loss function is defined as:
The total loss function $\mathcal{L}_{\text {total }}$ can be defined as:
\begin{equation}
\mathcal{L}_{\text {total }}=\mathcal{L}_{\text {task }}+\lambda_{\mathrm{MMD}} \cdot \operatorname{MMD}^2\left(X, X^{\prime}\right)
\end{equation}

\begin{align}
\operatorname{MMD}^2\left(X, X^{\prime}\right) &= \frac{1}{n^2} \sum_{i=1}^n \sum_{j=1}^n k\left(x_i, x_j\right) \nonumber \\
&\quad + \frac{1}{n^2} \sum_{i=1}^n \sum_{j=1}^n k\left(x_i^{\prime}, x_j^{\prime}\right) \nonumber \\
&\quad - \frac{2}{n^2} \sum_{i=1}^n \sum_{j=1}^n k\left(x_i, x_j^{\prime}\right)
\end{align}

$\mathcal{L}_{\text {task }}$ is the standard task-specific loss.
$\lambda_{\mathrm{MMD}}$ is a hyperparameter that balances the task loss and the MMD loss.
$X$ and $X^{\prime}$ are feature representations from different mini-batches or subsets within training data.

The model was trained for 100 epochs using the Adam optimizer with an initial learning rate of $10^{-3}$, weight decay of $1 \times 10^{-4}$, and the Noam learning rate schedule~\cite{vaswani2017attention}. The Noam schedule dynamically adjusts the learning rate based on the model’s hidden dimension(transformer embedding size) and training step, using the parameters $\beta_1=0.9$, $\beta_2=0.98$, and $\epsilon=10^{-9}$. Early stopping was applied with a patience of 7 epochs to prevent overfitting. All experiments were conducted on an RTX 4060 GPU.

\section{Results}
\begin{figure*}[t]
\centering
\begin{subfigure}[b]{.49\textwidth}
  \includegraphics[width=\linewidth]{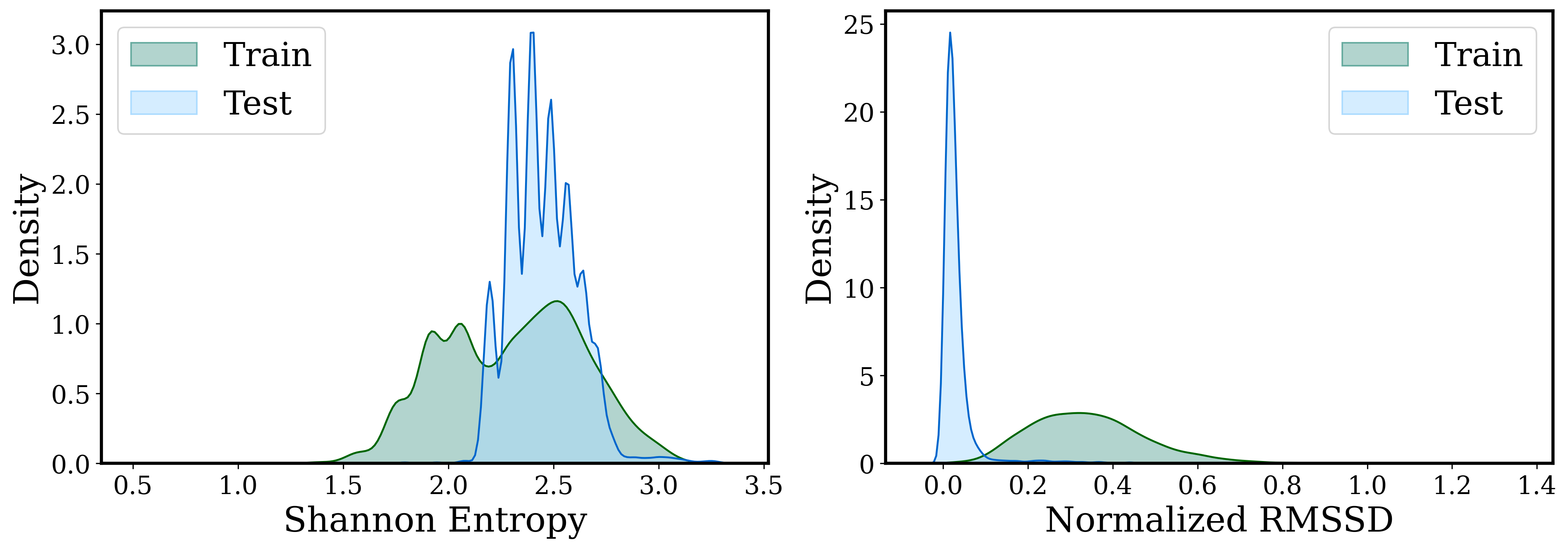}
  \caption{}
  \label{kde_a}
\end{subfigure}
\hfill
\begin{subfigure}[b]{.49\textwidth}
  \includegraphics[width=\linewidth]{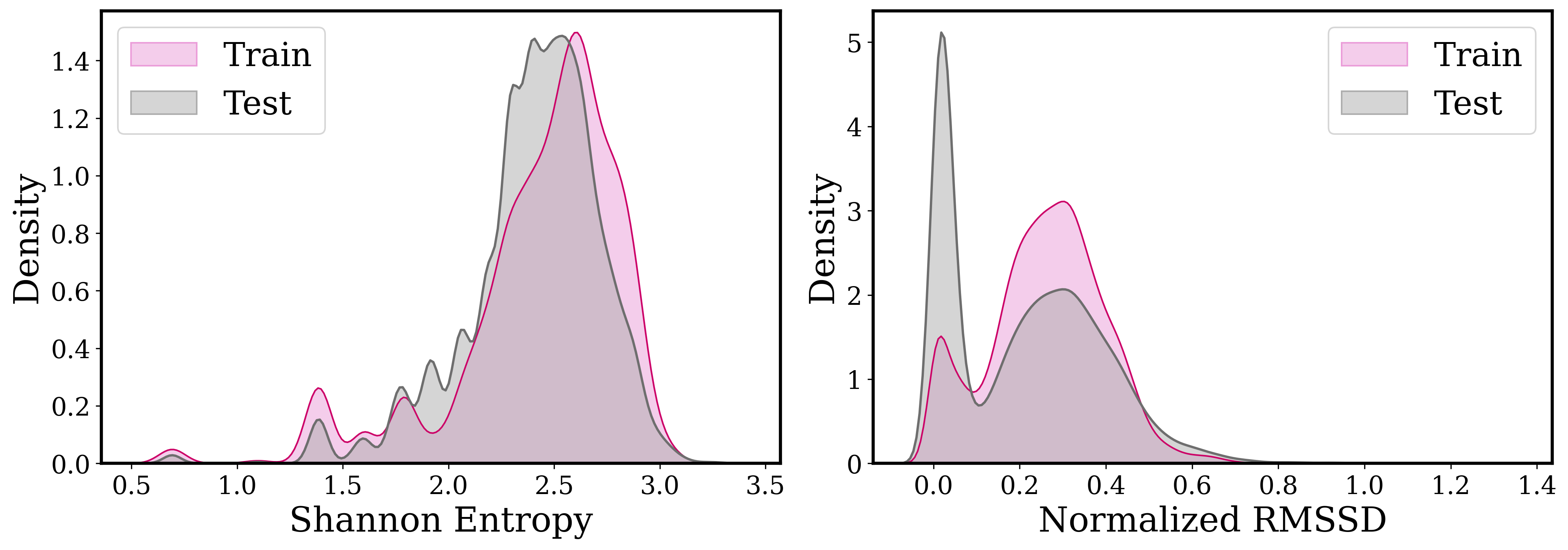}
  \caption{}
  \label{kde_b}
\end{subfigure}
\caption{KDE plots for Shannon Entropy and Normalized RMSSD before(a) and after preprocessing(b)}
\end{figure*}

\begin{figure*}[t]
\centering
\begin{subfigure}[b]{.99\textwidth}
  \includegraphics[width=\linewidth,height=3cm]{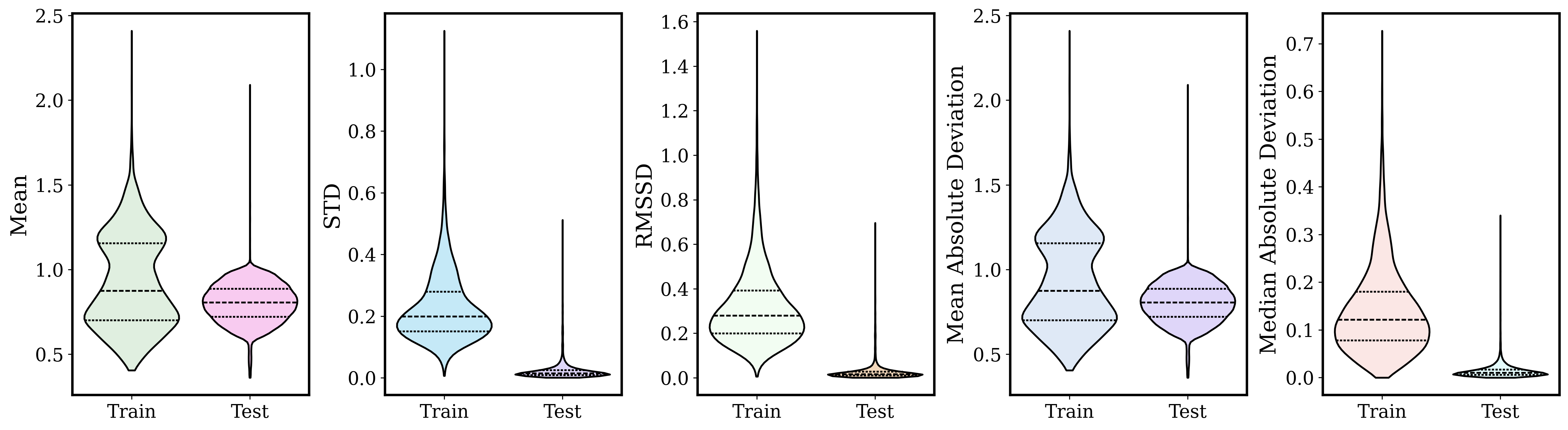}
  \caption{Violin plot before preprocessing.}
  \label{violin_a}
\end{subfigure}
\begin{subfigure}[b]{.99\textwidth}
  \includegraphics[width=\linewidth,height=3cm]{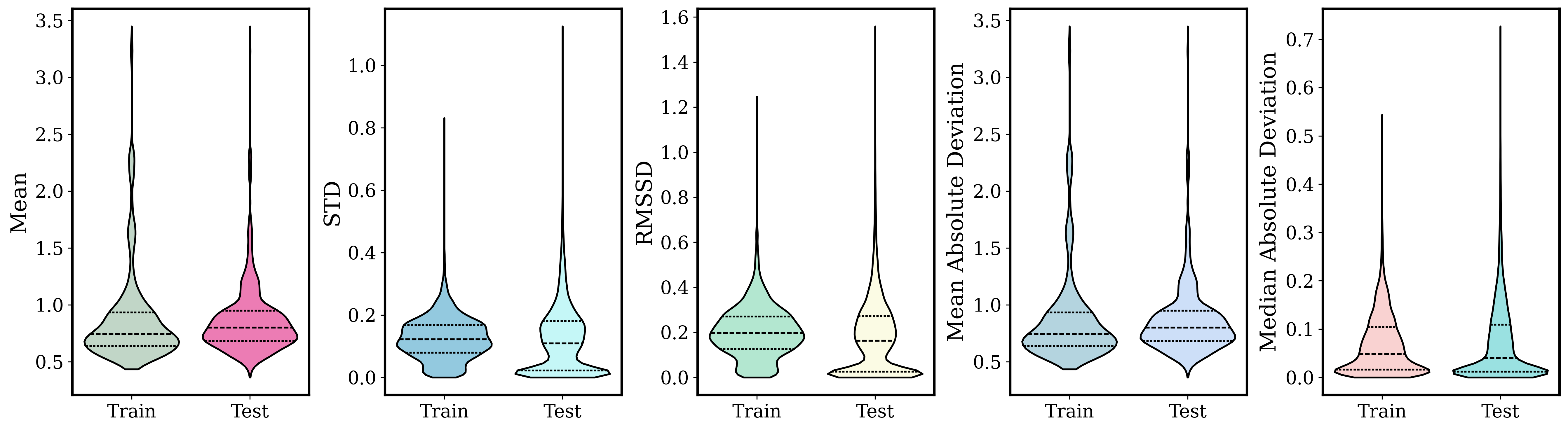}
  \caption{Violin plot after preprocessing.}
  \label{violin_b}
\end{subfigure}
\caption{Violin plots for Mean, Standard Deviation (STD), Root Mean Square of Successive Differences (RMSSD), Mean Absolute Deviation, and Median Absolute Deviation before (a) and after (b) preprocessing.}
\label{fig:violin_plots} 
\end{figure*}
The following sections present the performance of the proposed hybrid transformer model across different datasets and scenarios. To assess the effectiveness of the model, we first visualize the feature distributions across domains using Kernel Density Estimation (KDE) and violin plots. Subsequently, we evaluate the model's performance under three scenarios: training and testing on seen domains, generalization to an unseen domain, and the impact of excluding RR-interval features. Each scenario highlights an aspect of the model’s robustness and generalization.

\label{results}

\subsection{Visualization of Feature Distributions Across Domains}
We employed two visualization techniques to illustrate the domain differences in our ECG datasets: violin plots and KDE plots. KDE plots were prioritized for Shannon Entropy and Normalized RMSSD because these features often exhibit multi-modal or highly skewed distributions; KDE provides a smoother, more detailed view of these density shifts, which is critical for identifying subtle domain overlaps. Conversely, violin plots were selected for the remaining five statistical features because they effectively combine a box plot’s summary of central tendency and variance with a density curve, allowing for a direct side-by-side comparison of the ranges and quartiles between training and test sets.

 We utilized KDE plots for \textbf{Shannon Entropy} and \textbf{Normalized RMSSD} features. These features were chosen due to their ability to capture distinct characteristics of the ECG signals. The results highlight substantial domain shifts before and after data preprocessing.
 Before preprocessing, the KDE plots in Figure ~\ref{kde_a}, showed a noticeable difference in the distributions of Shannon Entropy and Normalized RMSSD between the training and test datasets. For Shannon Entropy, the test dataset exhibited a higher concentration of values between 2.0 and 2.5, while the training dataset was more spread out, suggesting different underlying data structures between the domains. In the case of Normalized RMSSD, the test dataset had a sharp peak at lower values, indicating a potential mismatch between the datasets regarding variability in heart rate signals.
After preprocessing, the KDE plots in Figure ~\ref{kde_b} demonstrated improved alignment between the training and test datasets. For Shannon Entropy, the distributions became more comparable, particularly in the 2.0 to 2.5 range, indicating a decrease in domain differences. Similarly, for Normalized RMSSD, the test dataset's distribution moved closer to the training dataset, although some differences remained.

Violin plots were used to compare the distribution of features such as \textbf{Mean}, \textbf{STD}, \textbf{RMSSD}, \textbf{Mean Absolute Deviation}, and \textbf{Median Absolute Deviation} between the training and test sets before and after preprocessing. These features represent basic statistical measures that describe central tendency and variability within the ECG signals.
In the preprocessed data in Figure~\ref{violin_a}, the test set consistently exhibited narrower distributions across all features, indicating a lack of variance compared to the training set. 
After preprocessing shown in Figure ~\ref{violin_b}, there was a noticeable alignment in the distributions between the training and test sets for most features. 
While Figure~\ref{violin_b} shows that some differences in the density peaks for STD and RMSSD still persist, the overall variability of the test set was brought significantly closer to the training set. These remaining discrepancies underscore the inherent challenge of domain shift in real-world wearable data and justify our use of Maximum Mean Discrepancy (MMD) during training to force the model to learn features that are invariant to these residual distribution gaps.

\begin{table*}[t]
    \centering
    \caption{Comparison of F1-score, Precision, and Recall for Normal, AF, and PVC across three scenarios.}
    \resizebox{\textwidth}{!}{  
    \begin{tabular}{lccccccccc}
        \toprule
         & \multicolumn{3}{c}{\textbf{Scenario I}} & \multicolumn{3}{c}{\textbf{Scenario II}} & \multicolumn{3}{c}{\textbf{Scenario III}} \\
        \cmidrule(r){2-4} \cmidrule(r){5-7} \cmidrule(r){8-10}
        Class & \textbf{Precision} & \textbf{Recall} & \textbf{F1-score} & \textbf{Precision} & \textbf{Recall} & \textbf{F1-score} & \textbf{Precision} & \textbf{Recall} & \textbf{F1-score} \\
        \midrule
        \textbf{Normal} &  98.74\% & 98.76\% & 98.8\% &97.78\%  & 91.17\% & \textbf{94.36\%} & 83.47\%  & 95.97\% & 89.29\% \\
        \textbf{AF} & 96.37\% & 95.72\% & 96.04\% & 85.97\% & 94.93\% & \textbf{90.23\%} &63.39\%  & 91.22\% & 74.8\% \\
        \textbf{PVC} &  98.05\% & 98.29\% & 98.17\% &95.71\%  & 95.43\%  & \textbf{95.57\%}  & 95.0\% & 91.41\% & 92.91\% \\
        \bottomrule
    \end{tabular}
    }
    \label{table_metrics}
\end{table*}

\subsection{Evaluation of model performance across training datasets without domain consideration}
In this scenario, we trained the model on five datasets and evaluated its performance on the same domains seen during training. The data was split into three subsets in a 60/20/20 ratio, with 60\% assigned for training, 20\% for validation, and 20\% for testing. This approach ensures that both the training and testing data come from the same domains, allowing us to evaluate the model's performance without the influence of domain shifts. The performance results are presented in Table~\ref{table_metrics}.
The model achieved high F1-scores, the harmonic mean of precision and recall, across all classes: 98.80\% for the Normal class, 96.04\% for AF, and 98.17\% for PVC. These scores indicate strong classification performance when the different domains are seen during training.

\subsection{Evaluation of model performance and generalization to unseen domain}

In this scenario, the model was evaluated on wearable device data, representing an unseen domain not learned during training. This analysis was critical for understanding the model’s ability to generalize to new data sources with different characteristics. As expected, there was a decrease in performance compared to the first scenario, where the test data was from domains previously seen by the model during training.
The results in Table~\ref{table_metrics} highlight the impact of domain shift on model performance. 

For the Normal class, the F1-score was 94\%, and for AF, it reached 90\%. Notably, the PVC class achieved a higher F1-score of 95\%, highlighting the effectiveness of the proposed model in detecting this arrhythmia type.


\begin{figure}[t] 
\centering
\begin{minipage}[t]{0.49\textwidth}
  \centering
  \includegraphics[width=7cm, height=5.3cm]{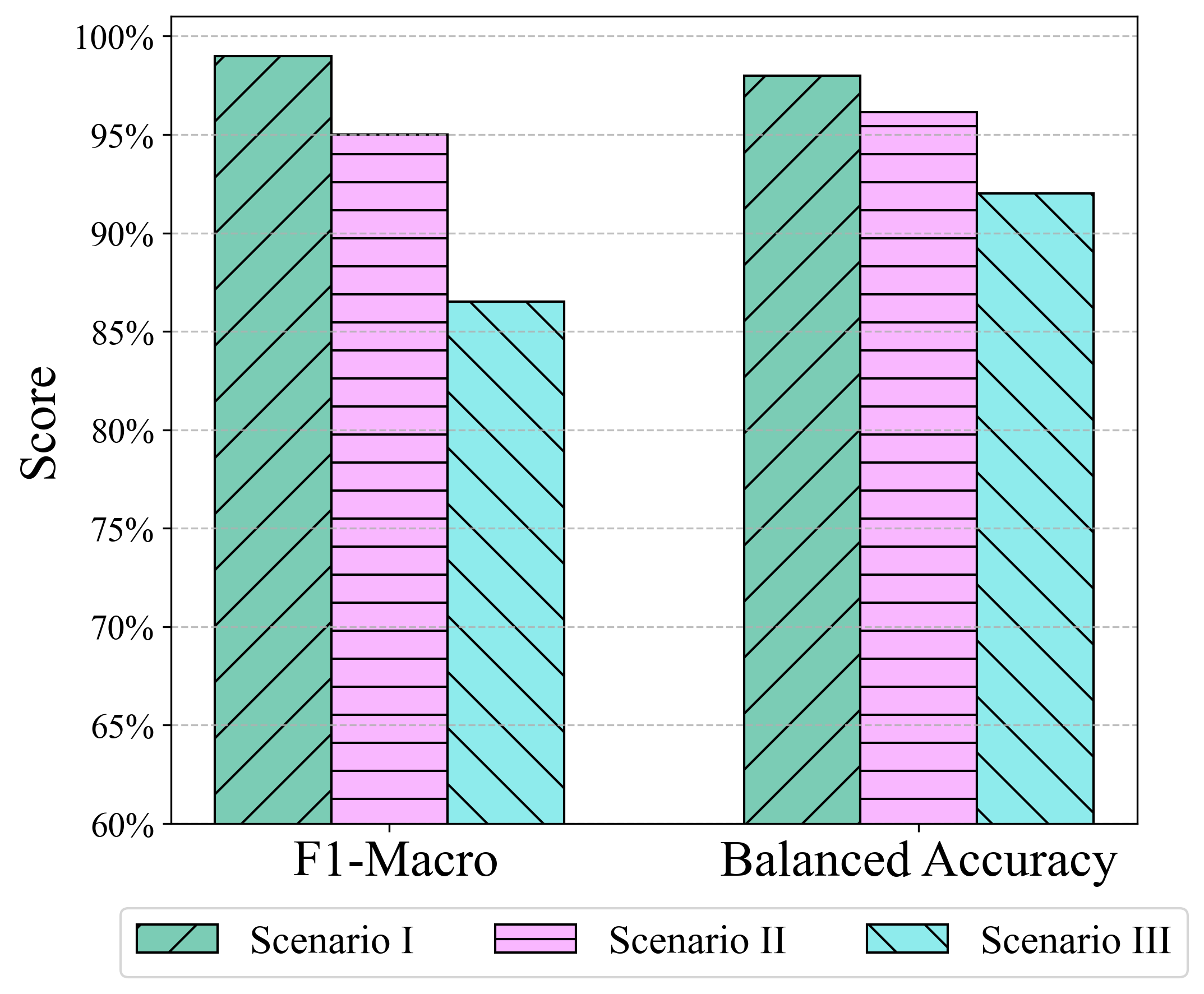}
  \caption{FF1-macro and balanced accuracy comparison across three scenarios (Scenario I, Scenario II, and Scenario III) for the Normal, AF, and PVC classes. Scenario I corresponds to testing on seen domains, Scenario II evaluates generalization to an unseen domain, and Scenario III assesses the model's performance without RR-interval features.}
  \label{fig:overal_comparison}
\end{minipage}%
\hfill
\begin{minipage}[t]{0.49\textwidth}
  \centering
  \includegraphics[width=7cm, height=5.2cm]{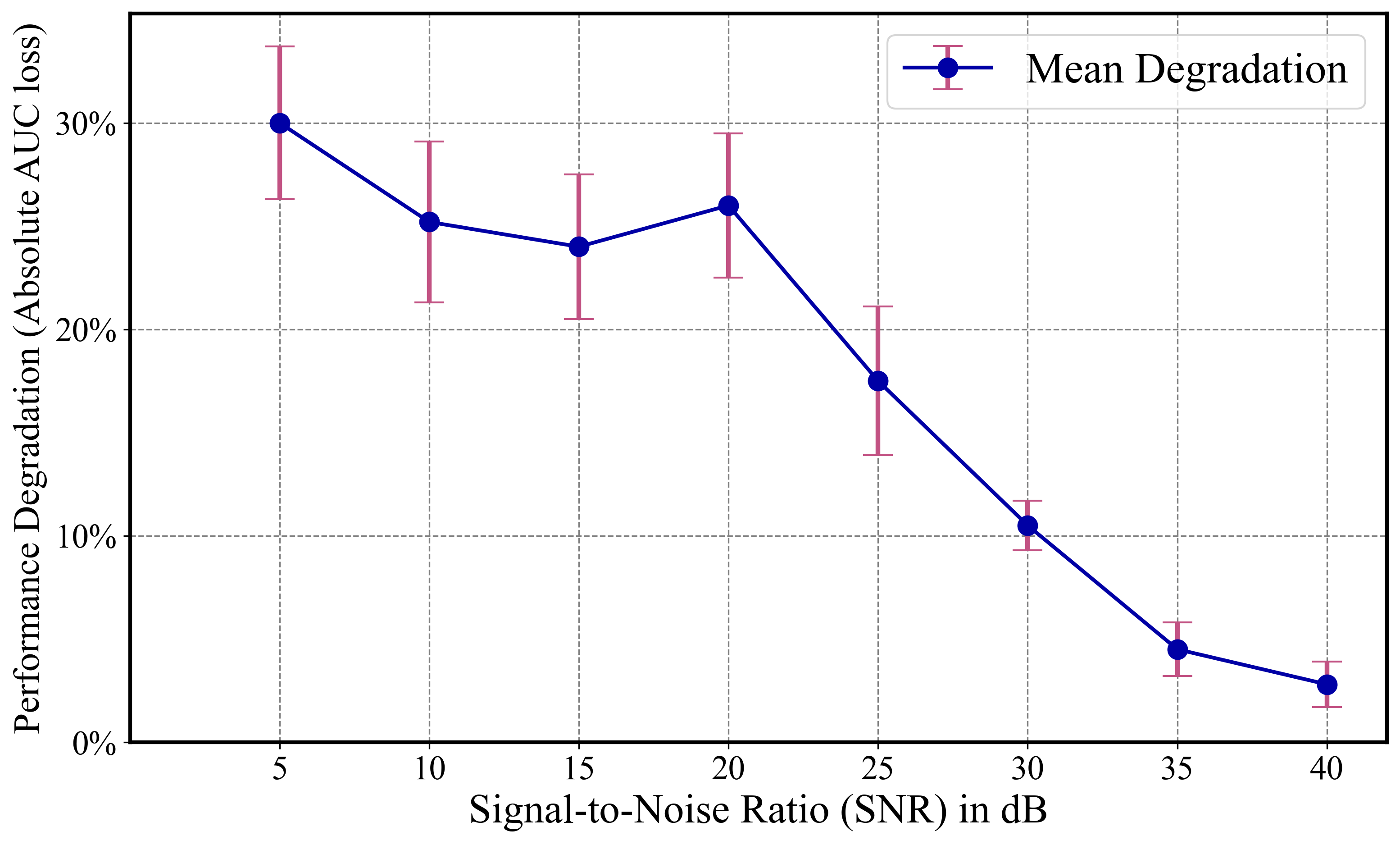}
  \caption{ Impact of SNR on model performance. The mean performance degradation (absolute AUC loss) is plotted against varying levels of additive white Gaussian noise. Error bars represent the standard deviation. The results show that the model is resilient to noise, with performance loss decreasing significantly as signal quality improves.}
  \label{fig:af_noise}
\end{minipage}
\end{figure}

\subsection{Effect of RR-Interval features on model performance}

In our last scenario, we investigated the impact of the RR-interval features on the model's performance. We retrained the model using only the 10-second ECG signals, excluding additional features from the RR intervals. Without these features, the model's performance dropped notably; for the Normal class, the F1-score dropped from 94\% in Scenario II to 89\% in Scenario III. For AF, the F1-score decreased significantly, from 90\% to 75\%. Similarly, for PVC, the F1-score decreases from 95\% to 93\%.

As shown in Figure~\ref{fig:overal_comparison}, which compares the three scenarios using overall metrics, we report  F1-macro and balanced accuracy to evaluate model performance. F1-macro assigns equal importance to each class, providing a comprehensive view of performance, while balanced accuracy measures the mean recall across classes, making both suitable for imbalanced multiclass evaluation.
As shown in Scenario II, the model demonstrates strong generalization to unseen wearable device data, achieving a high F1-macro 95\% and balanced accuracy 96.15\%, with only a slight drop of about 2\% compared to Scenario I. This highlights the effectiveness of the proposed model, where RR-interval features play a critical role, especially in handling complex arrhythmias. In Scenario III, removing these features leads to a more noticeable performance drop (e.g., a decrease of 8.5\% in F1-macro and 4\% in balanced accuracy), further confirming their value and supporting the model’s potential for real-world applications with variable data sources.

\subsection{Robustness to Noise}
To assess the robustness of our hybrid transformer model in realistic, non-controlled environments, we evaluated its performance on the wearable device test set after corrupting the signals with simulated noise. In practical scenarios, ECG signals from wearable devices are often distorted by noise stemming from motion artifacts, muscle activity, and poor electrode contact. To simulate these physiologically relevant conditions, we introduced additive white Gaussian noise at varying signal-to-noise ratios (SNRs) to investigate the impact of such real-world interferences on signal quality. This controlled corruption allows for a quantified assessment of the model’s stability under varying levels of signal degradation. 
 Figure~\ref{fig:af_noise} shows our model's robustness to noise, plotting the performance degradation (absolute AUC loss) against the SNR. A clear inverse relationship is evident: the model experiences its highest degradation of approximately 30\% at a low SNR of 5 dB. However, it demonstrates strong resilience as noise decreases, with the performance loss falling to under 5\% at an SNR of 35 dB. The minimal performance loss at higher SNRs demonstrates that the model maintains high structural stability and reliable classification, even when the input data is not pristine. This smooth degradation highlights the model's suitability for reliable performance in practical, noisy environments. Furthermore, the model’s ability to handle low-SNR inputs (5–15 dB) with predictable performance decay suggests it can maintain basic monitoring functionality in high-motion scenarios where traditional algorithms often fail


\section{Discussion} 
\begin{table*}[t]
\renewcommand{\arraystretch}{1.3}
\centering
\caption{Performance Comparison with Re-Implemented State-of-the-Art Models on Wearable ECG Data}
\label{tab:comparison_sota}

\rowcolors{4}{blue!7}{white}

\scriptsize   

\begin{tabular}{l c c l l c c}
\toprule
\multirow{2}{*}{\textbf{Reference}} &  \multirow{2}{*}{\textbf{Input Length}}  & \multirow{2}{*}{\textbf{\#Classes}} & \multirow{2}{*}{\textbf{Methodology}}  & \multirow{2}{*}{\textbf{Original Reported Metrics}}    & \multicolumn{2}{c}{\textbf{Performance on Wearable Data}} \\
\cmidrule(lr){6-7}
 &  &  &  &  & \textbf{F-macro} & \textbf{Balanced Accuracy} \\
\midrule
Jahan et al.~\cite{jahan2022short} & 20 heartbeats & 2 & HRV+AdaBoost  &  Accuracy: 88\%, F1 Scores: 87.99\%  & 78.66\%  & 80.1\% \\
Kim et al.~\cite{kim2022lightweight} & 10 seconds & 4 & Lightweight MobileNet  & Accuracy: 97.7\%, F1 Scores: 97.4\%  & 88.2\% & 92.8\%\\
Ribeiro et al.~\cite{ribeiro2020automatic} & 7-10 seconds & 6 &  ResNet  & F1 Scores: 87\%, Specificity:99\% &  86.98\% & 89.18\%   \\
Phukan et al.~\cite{phukan2023afibri} & 10 seconds & 2 &  5-layer 1D CNN & Accuracy: 99.84\%, Specificity: 93.23\% & 88.57\% & 90.9\%\\
Hassan et al.~\cite{hassan2022classification} & 3 seconds & 5 & CNN-BiLSTM &  Accuracy: 98\%, F1 Scores: 89.8\% &  85.62\% & 87.48\%  \\
Our method  & 10 seconds  & 3 &  CNN+Transformer & ------- & \textbf{95\%}  &\textbf{96.15\%} \\
\bottomrule
\end{tabular}
\end{table*}

\subsection{Comparison with State-of-the-Art Models}
To demonstrate the effectiveness of our proposed model, we selected a set of recent methods that achieved strong performance in arrhythmia detection and re-implemented their architectures within our experimental framework. The selected baselines include machine learning method, CNN and CNN-BiLSTM models that have demonstrated competitive results in prior studies. However, a direct comparison is not entirely fair due to key differences in training and test datasets, input signal length, number and type of arrhythmia classes, and evaluation protocols. Most notably, our model is evaluated on a proprietary single-lead wearable ECG dataset that is not publicly available and differs significantly from the benchmark datasets used in prior work. 
To ensure a systematic comparison, we did not rely on results reported in existing literature; instead, all baseline architectures were fully re-implemented and evaluated directly on our internal wearable ECG dataset.
Since all baseline models—including those re-implemented—were evaluated exclusively on our internal dataset, which varies from public ECG databases in signal quality, sampling rate, and patient demographics, the performance scores reported in Table\ref{tab:comparison_sota} should be interpreted as internal benchmarks rather than direct reproductions of previously published results. To achieve this, all comparative data were converted into a single-lead format, and all models were applied equally using the same pre-processing pipeline, ensuring that the performance improvements observed are a direct result of the model architecture rather than variations in data quality or lead configuration. This uniform evaluation framework allows us to quantify the significant performance leap of our hybrid Transformer, which achieved a 95\% F1-macro, over other re-implemented state-of-the-art methods.

\subsection{Future Work}
In future work, we aim to integrate an automated algorithm for detecting non-ECG regions to further enhance the quality and reliability of the input signals. This preprocessing step will be especially important for deployment in uncontrolled, real-world environments where noise and motion-induced distortions are common. Additionally, we aim to explore the real-time implementation of the proposed model on edge devices by optimizing the inference part. This would enable on-device arrhythmia detection suitable for wearable or mobile health monitoring systems.
\section{Conclusion}
This study proposes a hybrid transformer model that effectively combines raw ECG signals and HRV features to enhance arrhythmia detection, even in cross-domain settings. By incorporating domain adaptation techniques, the model addresses the critical challenge of domain shift, achieving high accuracy across both public datasets and unseen wearable device data. Despite minor performance degradation when applied to more complex arrhythmias like atrial fibrillation, the model demonstrates strong generalization capabilities. The results underscore the potential of integrating this approach into wearable ECG monitoring systems, which could improve real-time cardiovascular health tracking. Future work should focus on validating the model with larger, more diverse datasets to further improve its robustness and applicability in clinical settings, thus contributing to more accessible and effective cardiovascular care.
\section*{Data Availability}
The wearable data used for the test dataset in this study is publicly available and can be accessed at: \url{https://drive.google.com/file/d/1lF7rjGCYhr_8wA_lIUeG3iEJKy5oGICf/view?usp=sharing}

\bibliographystyle{elsarticle-num}   
\bibliography{references}



\end{document}